\documentclass[12pt]{article}
\pdfoutput=1
\usepackage{hyperref}
\usepackage{amsmath,amssymb,amsthm,amscd}
\usepackage{mathtools}
\usepackage{bm}
\usepackage{graphicx}

\newcommand{\dd}{\mathrm{d}}

\newcommand{\Tr}{\textrm{Tr}}

\newcommand{\be}{\begin{equation}}
\newcommand{\ee}{\end{equation}}
\newcommand{\ba}{\begin{aligned}}
\newcommand{\ea}{\end{aligned}}

\numberwithin{equation}{section}

\begin{document}
\hypersetup{pageanchor=false}
\begin{titlepage}

\rightline{USTC-ICTS/PCFT-26-72}

\vskip 3 cm

\begin{center}
{\Large\bfseries Giant Graviton Expansions\\[4pt]
from Symmetric Functions and Free Fermions\par}
\end{center}

\vskip 0.5 cm

\renewcommand{\thefootnote}{\fnsymbol{footnote}}
\vskip 30pt \centerline{ {\large \rm 
Min-xin Huang\footnote{minxin@ustc.edu.cn}  
} } \vskip .5cm  \vskip 20pt 

\begin{center}
{Interdisciplinary Center for Theoretical Study,  \\ \vskip 0.1cm  University of Science and Technology of China,  Hefei, Anhui 230026, China} 
 \\ \vskip 0.3 cm
{Peng Huanwu Center for Fundamental Theory,  \\ \vskip 0.1cm  Hefei, Anhui 230026, China} 
\end{center}

\setcounter{footnote}{0}
\renewcommand{\thefootnote}{\arabic{footnote}}
\vskip 40pt
\begin{abstract}

Giant graviton expansions express finite-rank corrections to supersymmetric
indices in terms of indices for wrapped D3-branes. We give detailed algebraic
proofs of two previously proposed relations for four-dimensional $\mathcal N=4$ $U(N)$
super-Yang--Mills theory. The first is the Gaiotto--Lee single-sum expansion
of the full superconformal index, proved by Hall--Littlewood orthogonality
and an elementary calculation of operator traces. The second is the
symmetric double-sum expansion of the flavored Schur index, proved by
factorizing its exact free-fermion mode product. We specify the analytic
continuation of the brane indices and
justify the sums coefficient by coefficient. The two proofs apply to
different expansion prescriptions and do not establish the full
three-species expansion of Imamura. We discuss the additional physical
and mathematical questions raised by that expansion and by the modular
properties of the Schur index.

\end{abstract}

\end{titlepage}
\hypersetup{pageanchor=true}
\vfill \eject


\newpage

\baselineskip=16pt

\tableofcontents

\section{Introduction}\label{sec:introduction}

The superconformal index gives an exact count, with bosonic and fermionic
signs, of protected states in a supersymmetric quantum field theory
\cite{Romelsberger:2005eg,Kinney:2005ej}. Its independence of continuous
couplings makes it a particularly useful observable in the AdS/CFT
correspondence \cite{Maldacena:1997re}. For four-dimensional
$\mathcal N=4$ $U(N)$ super-Yang--Mills theory, the large-$N$ index
has a simple interpretation in terms of Kaluza--Klein excitations of
type IIB supergravity on $AdS_5\times S^5$. At finite $N$, relations
among traces modify this spectrum. Giant gravitons, which are D3-branes
expanded into three-spheres inside $S^5$, provide a natural bulk
description of corrections whose charges grow with $N$
\cite{McGreevy:2000cw}. On the CFT side, sphere giant gravitons are
described by determinant and subdeterminant operators
\cite{Balasubramanian:2001nh}. Schur polynomials organize the half-BPS
operators and their exact finite-$N$ correlators, providing a
complementary description of giant graviton states \cite{Corley:2001zk}.
Understanding these corrections is a concrete
way to study how the finite number of colors is encoded in quantum
gravity.

The insightful proposal of Arai, Fujiwara, Imamura and Mori for the
Schur index \cite{Arai:2020qaj}, and Imamura's subsequent construction
for the full index \cite{Imamura:2021ytr}, turn this physical picture
into remarkably explicit relations. Write $I_N$ for the full index
and $I_\infty$ for its large-rank limit. In fugacities
$t,u,v,p,q$, which keep track of conserved charges and satisfy
$pq=tuv$, the proposed full expansion takes the form
\begin{equation}\label{eq:intro-imamura}
 \frac{I_N(t,u,v;p,q)}{I_\infty(t,u,v;p,q)}
 =\sum_{m_1,m_2,m_3\geq0}
 (t^{m_1}u^{m_2}v^{m_3})^N
 \mathcal W_{\boldsymbol m}(t,u,v;p,q).
\end{equation}
Here $\boldsymbol m=(m_1,m_2,m_3)$ lists the numbers of branes
wrapping the three types of cycles. The coefficient
$\mathcal W_{\boldsymbol m}$ is a worldvolume index for the coupled
$U(m_1)\times U(m_2)\times U(m_3)$ brane system, with the prescribed
integration cycles; the powers outside this coefficient are its
classical charge contribution. At generic fugacities the coefficients
are independent of $N$. The predictive strength of this construction
is that these coefficients are specified by brane dynamics, rather
than fitted to the finite-rank index.

Gaiotto and Lee developed an elegant complementary description in
terms of determinant operators and proposed a single-sum expansion
whose coefficients are smaller-rank indices with transformed
fugacities \cite{Gaiotto:2021xce}. This formulation singles out one
charge and uses an ordered expansion in the remaining fugacities.
Lee further developed the gauge-theory construction of brane indices
by counting modifications of determinant operators, including a
prescription for the brane integration contours \cite{Lee2022}.
Sameer Murthy supplied a powerful exact mathematical framework by
deriving a determinant expansion for a broad class of unitary matrix
integrals \cite{Murthy:2022ien}. Murthy's terms are explicitly defined,
but their identification with prescribed brane sectors is an
additional question: a term in that expansion need not correspond
to one fixed wrapping number. Liu and Rajappa made the arbitrary-order
determinant coefficients explicit and compared them with brane
coefficients \cite{Liu:2022olj}. Eniceicu further studied their
recombination into giant graviton contributions
\cite{Eniceicu:2023uvd}, and Chen, Mahajan and Tang obtained an
eigenvalue-instanton derivation of Murthy's expansion
\cite{Chen:2024cvf}. The analytic-continuation analysis of
Imamura \cite{Imamura:2022aua}, the comparison of symmetric and
asymmetric Schur expansions by Beccaria and Cabo-Bizet
\cite{Beccaria:2023zjw}, and the contour analysis of Ezroura, Liu and
Rajappa \cite{Ezroura:2024wmp} have substantially clarified this
distinction.

A residue-based proof of \eqref{eq:intro-imamura} would require more than finding
the expected worldvolume integrands at selected poles. One must
organize all residue configurations, including coincident poles,
and show that their sum gives precisely the physical integration
cycles of the three-node brane theory. Interactions involving all
three types of branes introduce cycle-dependent data that are absent
from a one-stack calculation. Moreover, analytic continuation can
change which poles contribute, so a change of fugacities cannot in
general be made while retaining the original contours. These are
structural issues, not just estimates of a remainder. We do not prove
the full three-species identity here. Instead, we treat two simpler
questions: the single-sum relation for the full index, and the
symmetric two-species relation in the Schur limit.
Special limits of the superconformal index, including the Schur,
Hall--Littlewood, Macdonald and Coulomb-branch limits, were studied
systematically in \cite{Gadde:2011uv}.

Several exact results make these questions accessible. Bourdier,
Drukker and Felix obtained the unflavored Schur index and developed
its free-fermion description
\cite{Bourdier:2015wda,Bourdier:2015sga}. The effective Fermi-gas
methods of Yasuyuki Hatsuda and Tadashi Okazaki give exact flavored indices and
provide the starting point for our second proof
\cite{Hatsuda:2022xdv}. Related finite-$N$ indices with antisymmetric
elliptic Ruijsenaars--Schneider surface-defect insertions admit a bilateral
free-fermion determinant representation on a special fugacity locus
\cite{Chae2026}. Yiwen Pan and Wolfger Peelaers derived closed expressions
for broad families of Schur indices using the ellipticity of the
integrand \cite{Pan:2021mrw}. Hatsuda's particularly useful
Macdonald-polynomial representation exposes the partition structure
of deformed Schur indices \cite{Hatsuda:2025mvj}; it motivates our
use of Hall--Littlewood polynomials for the first proof. A complementary
reformulation of unitary matrix models in terms of quantized symmetric
functions and a spin chain has also been developed
\cite{Ren:2026nrd}. Beccaria
and Cabo-Bizet obtained exact low-wrapping brane indices, described
their expansions near the unflavored limit by quasimodular forms,
and established the unflavored matching to all orders in the
fugacity through total wrapping number five
\cite[Section~4.2]{Beccaria:2024szi}. We keep the distinction between
all fugacity orders and all wrapping numbers explicit.

The Schur index is also attractive because of its modular structure.
The four-dimensional/two-dimensional correspondence of Beem,
Lemos, Liendo, Peelaers, Rastelli and van Rees identifies the protected
Schur sector with a two-dimensional chiral algebra
\cite{Beem:2013sza}. Its graded vacuum character, with the vacuum-energy
factor removed, is the Schur index. This gives a physical setting for
modular differential equations and high-temperature asymptotics,
as developed by Beem and Rastelli \cite{Beem:2017ooy} and by Pan and
Yang in explicit families and examples \cite{Pan:2024bne}.
Mixed-weight quasimodular expressions also make the structure of
unflavored indices particularly transparent \cite{Beem:2021zvt}.
In the $SU(N)$ theory, modular anomaly
equations and representations by generalized divisor sums provide
recursive control of the unflavored indices \cite{Huang:2022bry}.
Related methods extend to other gauge groups, where some of the
modular anomaly relations remain empirical \cite{Du:2023kfu}.
Here ``quasimodular'' means that the modular transformation law has
specified additional terms; it does not mean that the index is an
ordinary modular form of a single weight. The coexistence of this
structure with a sum over wrapped branes makes the Schur expansion
a useful setting in which to compare algebraic constraints and bulk
physics.

The main ingredient of our first proof is symmetric-function theory.
Hall--Littlewood orthogonality expresses the rank dependence of a matrix
integral through the lengths of partitions. A generating operator measures
these lengths, and an elementary trace calculation identifies every
coefficient with the required smaller-rank index. The operator and trace
methods build on the Macdonald free-field formalism
\cite{Feigin2009,Koshida2019} and the trace formula for periodic Macdonald
processes \cite{Koshida2021}.

Our second proof uses free fermions: in the exact Schur representation,
each integer mode is occupied at most once. We factor the generating product
into two generating series, each summed over all particle numbers.
Meromorphic continuation is applied to each fixed-particle-number coefficient
before we extract a Laurent coefficient.
The two particle numbers supply the two wrapping labels, while the
theta-function coefficients supply their interaction factor. Finite
coefficient counting ensures that each fixed radial order receives only
finitely many wrapping contributions and justifies the symmetric expansion.

Section~\ref{sec:proof} proves the full-index single-sum identity.
Section~\ref{sec:schur} proves the flavored Schur double-sum
identity. We state the standard symmetric-function
and free-fermion identities used as inputs and explain the subsequent
algebra and continuation in detail. Section~\ref{sec:conclusion}
discusses the remaining questions and their physical implications.

\section{Proof of the single-sum expansion}\label{sec:proof}

We prove the single-sum giant graviton expansion of the full
$\mathcal N=4$ $U(N)$ superconformal index.  The argument also proves a
more general identity for unitary matrix integrals.  We first define the
objects used in the proof.  We state the Hall--Littlewood Cauchy identity and its finite-variable
orthogonality formula explicitly as standard mathematical inputs.
For completeness, we derive the required specializations of the
established Macdonald operator and trace identities
\cite{Feigin2009,Koshida2019,Koshida2021}, together with the limiting steps.

\subsection{The index and the meaning of the expansion}

The group $U(N)$ consists of complex $N\times N$ matrices $U$ satisfying
$U^\dagger U=1$, where $U^\dagger$ is the conjugate transpose and $1$ is
the identity matrix.  We use the normalized Haar measure $\dd U$:
this is the invariant integration measure on $U(N)$, with
$\int_{U(N)}\dd U=1$.  The symbol $\Tr$ denotes the matrix trace, the sum
of diagonal entries.  The full index has the matrix-integral expression
\cite{Spiridonov:2010qv,Gaiotto:2021xce,Hatsuda:2025mvj}
\begin{equation}\label{eq:full-index}
 I_N(t,u,v;p,q)=\int_{U(N)}\dd U\,
 \exp\left[\sum_{n=1}^{\infty}\frac{f_n}{n}
             \Tr(U^n)\Tr(U^{-n})\right],
\end{equation}
where
\begin{equation}\label{eq:full-letter}
 f_n=1-\frac{(1-t^n)(1-u^n)(1-v^n)}{(1-p^n)(1-q^n)},
 \qquad pq=tuv.
\end{equation}
The variables $t,u,v,p,q$ are fugacities, meaning variables that keep
track of charges in the index.  The relation $pq=tuv$ leaves four
independent fugacities.  The phrase \emph{single-letter index} refers to
$f_1$; $f_n$ is obtained by raising every fugacity in $f_1$ to its $n$th
power.  We set $I_0=1$.

We use the finite and infinite products
\begin{equation}\label{eq:pochhammer-def}
 (a;b)_m=\prod_{j=0}^{m-1}(1-ab^j),\qquad
 (a;b)_0=1,\qquad
 (a;b)_\infty=\prod_{j=0}^{\infty}(1-ab^j).
\end{equation}
Here the finite length $m$ is a nonnegative integer. An empty product
is always $1$.  An ordinarily convergent infinite
product in \eqref{eq:pochhammer-def} will be used only when $|b|<1$;
elsewhere we will explicitly specify coefficientwise continuation.

The large-rank index and the expansion to be proved are
\begin{align}
 I_\infty(t,u,v;p,q)
 &=\frac{(p;p)_\infty(q;q)_\infty}
 {(t;t)_\infty(u;u)_\infty(v;v)_\infty},
 \label{eq:full-large-rank}\\
 \frac{I_N(t,u,v;p,q)}{I_\infty(t,u,v;p,q)}
 &=\sum_{k=0}^{\infty}t^{kN}
       I_k(t^{-1},q,p;v,u).
 \label{eq:full-gge}
\end{align}
Equation \eqref{eq:full-gge} is the proposal of Gaiotto and Lee
\cite{Gaiotto:2021xce}, in the conventions of
Eqs.~(3.27)--(3.30) of \cite{Hatsuda:2025mvj}.
The integer $N\geq0$ is the original rank, and the summation index
$k\geq0$ is the rank of the index on the right.

We prove \eqref{eq:full-gge} as an identity of formal series in the four
small fugacities $u,v,p,q$, subject to $pq=tuv$, with
$0<|t|<1$ kept fixed.
``Formal'' means that equalities are checked separately for each
coefficient; convergence of the complete series in the small fugacities
is not required.  Equivalently, put
\begin{equation}\label{eq:physical-scaling}
 (u,v,p,q)=\varepsilon(u_0,v_0,p_0,q_0),
 \qquad p_0q_0=tu_0v_0,
\end{equation}
and compare every coefficient of $\varepsilon$.  Here $\varepsilon$ is
a bookkeeping variable and $u_0,v_0,p_0,q_0$ are arbitrary symbolic
parameters subject to the displayed relation. The coefficient identities
hold for all these compatible parameters, rather than only along one
numerically chosen scaling ray.
This scaling preserves the balancing relation in
\eqref{eq:full-letter}.  In $I_k(t^{-1},q,p;v,u)$, each such coefficient
is evaluated first with the first fugacity of modulus less than one,
and then continued as a rational function to $t^{-1}$.  Keeping the
original integration contour after this substitution would in general
give a different prescription. During this construction, the same
matrix-integral expression is treated as an auxiliary function of five
arguments without imposing balance. Its coefficients are continued
rationally, and the balancing relation is imposed afterwards.
The continuation and the infinite
$k$-sum will be justified below.

\subsection{Partitions and Hall--Littlewood polynomials}

A \emph{partition} is a finite nonincreasing list of positive integers,
$\lambda=(\lambda_1,\ldots,\lambda_\ell)$.
Its length and size are
\begin{equation}
 \ell(\lambda)=\ell,\qquad
 |\lambda|=\lambda_1+\cdots+\lambda_\ell.
\end{equation}
We allow the empty partition $\varnothing$, with length and size zero.
Trailing zeros can be added whenever a list of a specified length is
needed.  For example, $(3,1)$ can be written $(3,1,0,0)$ in four
variables, but its partition length is still two.
The integer $m_r(\lambda)$ counts the parts equal to $r$.
For positive $r$ it does not depend on padding; in a list padded to
length $M$, $m_0(\lambda)=M-\ell(\lambda)$.

For $X=(x_1,\ldots,x_M)$, define its power sums by
\begin{equation}\label{eq:power-sums}
 p_n(X)=\sum_{a=1}^M x_a^n\qquad(n\geq1).
\end{equation}
The subscripted power sum $p_n(X)$ is distinct from the scalar
fugacity $p$.
A symmetric polynomial is a polynomial unchanged by permutations of
the $x_a$.  To discuss arbitrary numbers of variables at once, it is
convenient to regard the power sums $p_1,p_2,\ldots$ as independent
generators of the polynomial ring
\begin{equation}\label{eq:symmetric-ring}
 \Lambda=\mathbb C(t)[p_1,p_2,\ldots],\qquad \deg p_n=n.
\end{equation}
Here $\mathbb C(t)$ is the field of rational functions in $t$ with complex
coefficients.  Every polynomial in $\Lambda$ involves finitely many
$p_n$.  Evaluation in $M$ variables means substituting the finite sums
$p_n(X)$ for these generators, written $p_n\mapsto p_n(X)$.
The degree convention means, for example, that
$p_1^2$ and $p_2$ both have degree two.  A polynomial is homogeneous
of degree $d$ if each of its monomials has degree $d$.
The subspace $\Lambda_d$ of homogeneous degree $d$ is
finite-dimensional: its monomial basis is indexed by
partitions of $d$.

The Hall--Littlewood polynomials $P_\lambda(X;t)$ form a useful basis of
symmetric polynomials.  The following explicit definition fixes their
normalization.  For $\ell(\lambda)\leq M$, put
\begin{align}
 v_{\lambda,M}(t)
 &=\prod_{r\geq0}\prod_{j=1}^{m_r(\lambda)}
                         \frac{1-t^j}{1-t},
 \label{eq:hl-normalizer}\\
 P_\lambda(X;t)
 &=\frac1{v_{\lambda,M}(t)}
 \sum_{\sigma\in S_M}\sigma\left[
 x_1^{\lambda_1}\cdots x_M^{\lambda_M}
 \prod_{1\leq a<b\leq M}\frac{x_a-tx_b}{x_a-x_b}
 \right].
 \label{eq:hl-definition}
\end{align}
The symbol $S_M$ is the set of all $M!$ permutations of $M$ objects,
and $\sigma$ permutes all the variables in the expression in brackets.
Although individual summands have denominators, their apparent poles
at $x_a=x_b$ cancel in the sum; the result is a polynomial.
We put $P_\lambda(X;t)=0$ when $\ell(\lambda)>M$ and
$P_\varnothing=1$.
These polynomials are homogeneous of degree $|\lambda|$ and are stable
under setting the last variable to zero.  Thus they define elements of
$\Lambda$, also denoted by $P_\lambda$.
For example,
\begin{equation}\label{eq:hl-examples}
 P_{(1)}=p_1,\qquad
 P_{(1,1)}=\frac{p_1^2-p_2}{2},\qquad
 P_{(2)}=\frac{(1-t)p_1^2+(1+t)p_2}{2}.
\end{equation}
Their basis property lets us use ordinary linear algebra, separately
in each finite-dimensional space $\Lambda_d$.

We need two standard identities for these polynomials
\cite[Chapters III and VI]{Macdonald1995}.
Define
\begin{equation}\label{eq:b-lambda}
 b_\lambda(t)=\prod_{r\geq1}(t;t)_{m_r(\lambda)}.
\end{equation}
The first identity is the Cauchy identity,
\begin{equation}\label{eq:hl-cauchy}
 \exp\left[\sum_{n\geq1}\frac{1-t^n}{n}
                            p_n(X)p_n(Y)\right]
 =\sum_\lambda b_\lambda(t)P_\lambda(X;t)P_\lambda(Y;t).
\end{equation}
The variables $Y$ are a second independent list of variables.
This identity is a generating-function equality: its two sides have
the same coefficients in $X,Y$.  It remains true with independent
power-sum symbols on either side, so we may apply a linear operator to
one set of power sums.

The second identity is finite-variable orthogonality, in the following
normalization:
\begin{align}
 &\int_{U(N)}\dd U\,
 \exp\left[\sum_{n\geq1}\frac{t^n}{n}
                   \Tr(U^n)\Tr(U^{-n})\right]
 P_\lambda(U;t)P_\mu(U^{-1};t)
 \nonumber\\[-2pt]
 &\hspace{2cm}=
 \frac{\delta_{\lambda\mu}}
 {b_\lambda(t)(t;t)_{N-\ell(\lambda)}}
 \qquad\bigl(\ell(\lambda),\ell(\mu)\leq N\bigr).
 \label{eq:hl-norm}
\end{align}
Here $P_\lambda(U;t)$ means evaluation at the eigenvalues of $U$,
and the Kronecker symbol $\delta_{\lambda\mu}$ is $1$ when the partitions
are identical and $0$ otherwise; the same convention applies to
$\delta_{nm}$ for integer labels. Orthogonality means precisely this
vanishing of the mixed integrals.  No assumption that the parameters
or polynomial coefficients are complex-conjugated is implicit in
\eqref{eq:hl-norm}.

We record how the measure in this formula is normalized, since the
diagonal factors are important.  For any expression $A(z)$ in
$z=(z_1,\ldots,z_N)$, write $\operatorname{CT}_z A$ for the coefficient
of $z_1^0\cdots z_N^0$ in its Laurent expansion.
When the expansion converges on unit circles, this is
\begin{equation}\label{eq:ct-integral}
 \operatorname{CT}_z A
 =\oint\cdots\oint\prod_{i=1}^N\frac{\dd z_i}{2\pi\mathrm{i}z_i}
 A(z),\qquad \mathrm{i}^2=-1,
\end{equation}
with all circles traversed counterclockwise.
Weyl's integration formula for symmetric functions of unitary
eigenvalues is
\begin{equation}\label{eq:weyl}
 \int_{U(N)}\dd U\,A(U)
 =\frac1{N!}\operatorname{CT}_z
 \left[\prod_{i\ne j}(1-z_i/z_j)A(z)\right].
\end{equation}
Using $\exp(\sum_{n\geq1}y^n/n)=(1-y)^{-1}$, we see that the
integral on the left of \eqref{eq:hl-norm} has the measure
\begin{equation}\label{eq:hl-measure}
 \frac1{N!(1-t)^N}\operatorname{CT}_z\Delta_t(z),
 \qquad
 \Delta_t(z)=\prod_{i\ne j}\frac{1-z_i/z_j}{1-tz_i/z_j}.
\end{equation}
The $N$ factors with $i=j$ give $(1-t)^{-N}$.  Formula
\eqref{eq:hl-norm} is the usual Hall--Littlewood norm with these
diagonal factors included.  It is also the specialization of the
Macdonald norm in \cite{Hatsuda:2025mvj} at zero value of its first
polynomial deformation parameter; this does not set the physical
fugacity $q$ to zero.

\subsection{A general matrix integral and its finite-rank formula}

For a sequence $f=(f_1,f_2,\ldots)$, define
\begin{equation}\label{eq:general-index}
 Z_N[f]=\int_{U(N)}\dd U\,
 \exp\left[\sum_{n\geq1}\frac{f_n}{n}
                  \Tr(U^n)\Tr(U^{-n})\right],\qquad Z_0[f]=1.
\end{equation}
We separate one bosonic contribution by writing
\begin{equation}\label{eq:g-def}
 f_n=t^n+(1-t^n)g_n.
\end{equation}
Initially the $g_n$ are independent formal variables of weight $n$.
The weight of $g_1^{a_1}\cdots g_m^{a_m}$ is
$\sum_{n=1}^m n a_n$.  At a specified weight there are finitely many
such monomials.  Later the $g_n$ may be replaced by any series with
$g_n=O(\varepsilon^n)$; this notation means that all powers of
$\varepsilon$ below $n$ have zero coefficients.  This convention makes
all intermediate computations finite at each requested order.

When the variables $g_n$ occur, we allow coefficients in $\Lambda$ to
be formal series in the $g_n$, interpreted weight by weight.
Define an operator $\mathcal T$ on this extended space by
\begin{equation}\label{eq:T-def}
 (\mathcal T F)(p_1,p_2,\ldots)
 =F(g_1p_1,g_2p_2,\ldots).
\end{equation}
It is linear and preserves degree.  Its matrix in the
Hall--Littlewood basis is defined by
\begin{equation}\label{eq:T-matrix}
 \mathcal T P_\lambda
 =\sum_{|\mu|=|\lambda|}T_{\mu\lambda}P_\mu,
 \qquad a_\lambda=T_{\lambda\lambda}.
\end{equation}
Every $T_{\mu\lambda}$ has weight $|\lambda|$ in the $g_n$ and is
rational in $t$.  In particular $a_\varnothing=1$.

The trace of a linear operator on a finite-dimensional space is the
sum of its diagonal matrix entries; it is independent of the chosen
basis.  Here we take that trace on each $\Lambda_d$, then sum over
$d\geq0$.  This is a formal graded trace: at each weight in the $g_n$
only one finite-dimensional degree block contributes to
$\operatorname{Tr}_{\Lambda}\mathcal T$.
In the monomial basis $\prod p_n^{a_n}$ the operator $\mathcal T$ is
diagonal, with eigenvalue $\prod g_n^{a_n}$.  Geometric summation gives
\begin{equation}\label{eq:R-trace}
 R:=\operatorname{Tr}_{\Lambda}\mathcal T
 =\sum_\lambda a_\lambda
 =\prod_{n\geq1}\sum_{a=0}^{\infty}g_n^a
 =\prod_{n\geq1}\frac1{1-g_n}.
\end{equation}
The constant term of $R$ is $1$, so its inverse $R^{-1}$ is well-defined
as a formal series.

Apply $\mathcal T$ to the $Y$ power sums in \eqref{eq:hl-cauchy}.
Its left side becomes
\begin{equation}\label{eq:deformed-cauchy}
 \exp\left[\sum_{n\geq1}\frac{(1-t^n)g_n}{n}
                                   p_n(X)p_n(Y)\right]
 =\sum_\lambda b_\lambda(t)P_\lambda(X;t)
                                    (\mathcal T P_\lambda)(Y;t).
\end{equation}
Now set $X$ equal to the eigenvalues of $U$, set $Y=X^{-1}$, multiply
by the pure-$t$ exponential in \eqref{eq:hl-norm}, and integrate.
Here $X^{-1}=(x_1^{-1},\ldots,x_N^{-1})$.
By \eqref{eq:T-matrix}, orthogonality retains only the term
$\mu=\lambda$.  The factor $b_\lambda$ cancels the same factor in the
norm denominator.  We obtain
\begin{equation}\label{eq:finite-rank-HL}
 \boxed{\displaystyle
 Z_N[f]=\sum_{\ell(\lambda)\leq N}
                 \frac{a_\lambda}{(t;t)_{N-\ell(\lambda)}}.}
\end{equation}
Polynomials indexed by longer partitions vanish upon evaluation in
$N$ variables; they do not contribute.  At fixed weight only finitely
many partitions remain, so the expansion and integration commute
coefficientwise.

For $0<|t|<1$, take $N\to\infty$ in each coefficient of
\eqref{eq:finite-rank-HL}.  Using \eqref{eq:R-trace} gives
\begin{equation}\label{eq:general-large-rank}
 Z_\infty[f]=\frac{R}{(t;t)_\infty}
 =\prod_{n\geq1}\frac1{(1-t^n)(1-g_n)}
 =\prod_{n\geq1}\frac1{1-f_n}.
\end{equation}

\subsection{Extracting powers of the rank variable}

Let $\xi$ be an auxiliary rank variable, unrelated to any fugacity or
eigenvalue, and
define
\begin{equation}\label{eq:interpolation}
 \Phi(\xi)=\frac1R\sum_\lambda a_\lambda
                         (\xi t^{1-\ell(\lambda)};t)_\infty.
\end{equation}
For each fixed monomial in the $g_n$, its coefficient is a finite
linear combination of the products appearing here.  Each product is
an entire function of $\xi$ when $0<|t|<1$: its factors have the form
$1-c\xi t^j$ for a fixed $c$, and the sum of $|c\xi t^j|$ converges uniformly
for $\xi$ in every bounded set.  Thus the coefficient is defined for all
complex $\xi$ and equals its convergent Taylor series at $\xi=0$.

At $\xi=t^N$, a partition of length $\ell>N$ gives zero, since its
product contains the factor $1-t^{N+1-\ell+j}$ with
$j=\ell-N-1$, namely $1-1$.  For $\ell\leq N$,
\begin{equation}
 (t^{N+1-\ell};t)_\infty
 =\frac{(t;t)_\infty}{(t;t)_{N-\ell}}.
\end{equation}
Equations \eqref{eq:finite-rank-HL}--\eqref{eq:interpolation} therefore
imply
\begin{equation}\label{eq:interpolation-at-N}
 \Phi(t^N)=\frac{Z_N[f]}{Z_\infty[f]}.
\end{equation}

We next compute the Taylor coefficients of $\Phi$.
The elementary Euler expansion is
\begin{equation}\label{eq:euler-series}
 (\xi t;t)_\infty
 =\sum_{k=0}^{\infty}
 \frac{(-1)^kt^{k(k+1)/2}}{(t;t)_k}\xi^k
 =\sum_{k=0}^{\infty}\frac{\xi^k}{(s;s)_k},
 \qquad s=t^{-1}.
\end{equation}
For completeness, if $E(\xi)=(\xi t;t)_\infty=\sum c_k\xi^k$, then
$E(\xi)=(1-\xi t)E(t\xi)$ and $c_0=1$.  Comparing coefficients gives
$(1-t^k)c_k=-t^kc_{k-1}$, which proves the first equality.  The second
uses the finite-product identity
\begin{equation}\label{eq:poch-inversion}
 (t^{-1};t^{-1})_k=(-1)^kt^{-k(k+1)/2}(t;t)_k.
\end{equation}
No infinite product of base $t^{-1}$ has been used.
Replacing $\xi$ by $\xi s^\ell$ in \eqref{eq:euler-series} yields
\begin{equation}\label{eq:rank-coefficients}
 \Phi(\xi)=\sum_{k=0}^{\infty}\xi^k C_k,\qquad
 C_k=\frac1{(s;s)_kR}\sum_\lambda a_\lambda s^{k\ell(\lambda)}.
\end{equation}
We now identify $C_k$ as a rank-$k$ matrix integral.

\subsection{An operator that measures the length of a partition}

On the polynomial ring \eqref{eq:symmetric-ring}, let
$\partial_{p_n}$ be ordinary differentiation with respect to $p_n$.
Multiplication by $p_n$ and differentiation obey
$\partial_{p_n}p_m-p_m\partial_{p_n}=\delta_{nm}$.
Define the formal generating operator
\begin{equation}\label{eq:eta-definition}
 \eta(z)=
 \exp\left[\sum_{n\geq1}\frac{1-s^n}{n}p_nz^n\right]
 \exp\left[-\sum_{n\geq1}z^{-n}\partial_{p_n}\right],
 \qquad H=\operatorname{CT}_z\eta(z).
\end{equation}
Operators act from right to left.  The first exponential multiplies a
polynomial by a generating series; the second is a translation, because
Taylor's formula gives
\begin{equation}\label{eq:eta-shift}
 \exp\left[-\sum_{n\geq1}z^{-n}\partial_{p_n}\right]F(p_1,p_2,\ldots)
 =F(p_1-z^{-1},p_2-z^{-2},\ldots).
\end{equation}
For a polynomial of degree $d$, the right side contains only the powers
$z^0,z^{-1},\ldots,z^{-d}$.  Consequently only finitely many terms of
the multiplication exponential can contribute to its constant term.
Thus $H$ is a well-defined degree-preserving linear operator.

We will prove
\begin{equation}\label{eq:H-spectrum}
 \boxed{H P_\lambda=s^{\ell(\lambda)}P_\lambda.}
\end{equation}
The operator $\eta(z)$ and its constant term are the Hall--Littlewood
specialization of the Macdonald free-field realization; see
\cite[Proposition~3.6]{Feigin2009} and
\cite[Section~3]{Koshida2019}. We derive their action here in our
normalization, without assuming familiarity with vertex operators.

Evaluate in $M$ variables $X=(x_1,\ldots,x_M)$.  The multiplication
factor in \eqref{eq:eta-definition} becomes
\begin{equation}\label{eq:eta-rational}
 \exp\left[\sum_{n\geq1}\frac{1-s^n}{n}p_n(X)z^n\right]
 =\prod_{a=1}^M\frac{1-sx_az}{1-x_az}
 =s^M+\sum_{a=1}^M\frac{A_a(X)}{1-x_az},
\end{equation}
where partial fractions give
\begin{equation}
 A_a(X)=(1-s)\prod_{b\ne a}\frac{x_a-sx_b}{x_a-x_b}.
\end{equation}
These formulas can first be checked for distinct nonzero $x_a$; after
the denominators cancel, they hold as polynomial identities in general.
If $G(z^{-1})$ is a polynomial in $z^{-1}$, expansion of the geometric
series shows
\begin{equation}
 \operatorname{CT}_z\frac{G(z^{-1})}{1-x_az}=G(x_a).
\end{equation}
Apply this to the translated polynomial in \eqref{eq:eta-shift}.
Substitution $z^{-1}=x_a$ replaces each $p_n(X)$ by
$p_n(X)-x_a^n$, which is the power sum of the list with $x_a$ deleted.
Writing that list as $X\setminus x_a$, we find
\begin{equation}\label{eq:H-finite}
 HF(X)=s^MF(X)+(1-s)\sum_{a=1}^M
 \prod_{b\ne a}\frac{x_a-sx_b}{x_a-x_b}F(X\setminus x_a).
\end{equation}

Introduce the finite-variable operator
\begin{equation}\label{eq:D-definition}
 D_{M,t}F(X)=\sum_{a=1}^M
 \prod_{b\ne a}\frac{tx_a-x_b}{x_a-x_b}F(X\setminus x_a).
\end{equation}
Since $s=t^{-1}$, \eqref{eq:H-finite} says
\begin{equation}\label{eq:H-D}
 H=s^M\bigl[1+(t-1)D_{M,t}\bigr].
\end{equation}
We can compute $D_{M,t}P_\lambda$ directly from
\eqref{eq:hl-definition}.  Let $\ell=\ell(\lambda)<M$.
In each summand with $x_a$ deleted, insert the symmetrized formula for
$P_\lambda$ in $M-1$ variables.  Regard $x_a$ as the final variable,
with exponent zero.  The missing factors involving this final variable
are exactly
\begin{equation}
 \prod_{b\ne a}\frac{x_b-tx_a}{x_b-x_a}
 =\prod_{b\ne a}\frac{tx_a-x_b}{x_a-x_b}.
\end{equation}
Summing over $a$ and the permutations of the other variables therefore
produces the full $M$-variable symmetrization.  Only its normalization
changes, so
\begin{equation}\label{eq:D-spectrum}
 D_{M,t}P_\lambda
 =\frac{v_{\lambda,M}(t)}{v_{\lambda,M-1}(t)}P_\lambda
 =\frac{1-t^{M-\ell}}{1-t}P_\lambda.
\end{equation}
The last ratio comes solely from the extra zero part.  If $\ell=M$,
every evaluation in $M-1$ variables vanishes, and both sides of
\eqref{eq:D-spectrum} are zero.  Substitution in \eqref{eq:H-D} gives
\begin{equation}
 HP_\lambda
 =s^M\left[1+(t-1)\frac{1-t^{M-\ell}}{1-t}\right]P_\lambda
 =s^\ell P_\lambda,
\end{equation}
as required. On $\Lambda_d$, evaluation in $M\ge d$ variables is
injective: a homogeneous symmetric function of degree $d$ is determined
by such an evaluation, since all partitions of $d$ have length at most
$d$. Thus the finite-variable identity proves the identity in $\Lambda$.  For example, $HP_{(2)}=sP_{(2)}$, whereas
$HP_{(1,1)}=s^2P_{(1,1)}$: it is the number of nonzero rows, not the
total size, that is measured.

In the Hall--Littlewood basis, \eqref{eq:H-spectrum} and
\eqref{eq:T-matrix} now give
\begin{equation}\label{eq:C-as-trace}
 C_k=\frac{\operatorname{Tr}_{\Lambda}(\mathcal T H^k)}
 {(s;s)_kR}.
\end{equation}
No commutativity of $\mathcal T$ and $H$ is asserted or needed:
$H^k$ acts diagonally on the input basis vector, and the trace then
selects the diagonal coefficient of $\mathcal T$.

\subsection{Products of the generating operator}

For a product of several $\eta$ operators, \emph{normal ordering},
denoted by colons, means that all multiplication operators are placed
to the left of all differentiation operators, without retaining their
commutators inside the colons.  For $z=(z_1,\ldots,z_k)$ this means
\begin{align}\label{eq:normal-product}
 :\eta(z_1)\cdots\eta(z_k):
 ={}&\exp\left[\sum_{n\geq1}\frac{1-s^n}{n}p_np_n(z)\right]
 \exp\left[-\sum_{n\geq1}p_n(z^{-1})\partial_{p_n}\right],
\end{align}
where $p_n(z)=\sum_{a=1}^kz_a^n$ and
$p_n(z^{-1})=\sum_{a=1}^kz_a^{-n}$.

Taylor translation \eqref{eq:eta-shift} gives the scalar factor needed
when moving a differentiation exponential past a multiplication
exponential:
\begin{align}
 \eta(z)\eta(w)
 &=\exp\left[-\sum_{n\geq1}\frac{1-s^n}{n}(w/z)^n\right]
                         :\eta(z)\eta(w):\nonumber\\
 &=K_s(w/z):\eta(z)\eta(w):,
 \qquad K_s(y)=\frac{1-y}{1-sy}.
 \label{eq:eta-product}
\end{align}
The last equality follows from $-\sum_{n\geq1}y^n/n=\log(1-y)$.
Repeating this operation for every pair gives
\begin{equation}\label{eq:eta-k-product}
 \eta(z_1)\cdots\eta(z_k)
 =\prod_{i<j}K_s(z_j/z_i):\eta(z_1)\cdots\eta(z_k):.
\end{equation}

We need the rational-function identity
\begin{equation}\label{eq:symmetrization}
 \sum_{\sigma\in S_k}\prod_{i<j}
 K_s(z_{\sigma(j)}/z_{\sigma(i)})
 =\frac{(s;s)_k}{(1-s)^k}\Delta_s(z).
\end{equation}
Here $\Delta_s$ is defined by \eqref{eq:hl-measure} with $N=k$ and
$t$ replaced by $s$.  To prove this identity, divide its left side by
$\Delta_s(z)$.  The result is
\begin{equation}\label{eq:sym-quotient}
 \sum_{\sigma\in S_k}\prod_{i<j}
 \frac{z_{\sigma(j)}-s z_{\sigma(i)}}
      {z_{\sigma(j)}-z_{\sigma(i)}}.
\end{equation}
Its only possible poles are at equal variables.  Pair permutations
that interchange those two variables: their residues have opposite
signs, while the other factors agree when the variables coincide.
The poles therefore cancel.  The sum is a polynomial unchanged by
rescaling all variables together; since every term has degree zero,
that polynomial must be a constant.

Evaluate it in the limit $|z_1|\ll|z_2|\ll\cdots\ll|z_k|$.
Each factor tends to $1$ or $s$.  The exponent of $s$ is the inversion
number of the permutation, that is, the number of pairs $i<j$ for which
$\sigma(i)>\sigma(j)$.  Inserting the largest entry into a permutation
of $k-1$ entries creates $0,1,\ldots,k-1$ new inversions.  Induction
therefore gives
\begin{equation}
 \sum_{\sigma\in S_k}s^{\operatorname{inv}(\sigma)}
 =\prod_{j=1}^k(1+s+\cdots+s^{j-1})
 =\frac{(s;s)_k}{(1-s)^k},
\end{equation}
which proves \eqref{eq:symmetrization}.

The normally ordered product \eqref{eq:normal-product} is symmetric
in the $z_a$.  Constant-term extraction is also unchanged by renaming
these variables.  We can thus average the kernel of
\eqref{eq:eta-k-product} over all $k!$ permutations and use
\eqref{eq:symmetrization}.  Since taking a separate constant term from
each factor on the left of \eqref{eq:eta-k-product} gives $H^k$, we obtain
\begin{equation}\label{eq:H-power}
 \boxed{\displaystyle
 \frac{H^k}{(s;s)_k}
 =\frac1{k!(1-s)^k}\operatorname{CT}_z
 \left[\Delta_s(z):\eta(z_1)\cdots\eta(z_k):\right].}
\end{equation}
The conventions for $k=0$ make both sides the identity operator.
Formula \eqref{eq:H-power} is the Hall--Littlewood specialization,
in the present normalization, of the higher Macdonald-operator
formula in \cite[Proposition~3.6]{Feigin2009}, also stated in
\cite[Theorem~3.11]{Koshida2019}. The derivation above makes the
constant-term prescription explicit.

For precision, this computation may first be performed for $|s|<1$,
on matrix elements between fixed homogeneous polynomials.  A matrix
element means the coefficient of a chosen output basis polynomial
when the operator acts on a chosen input basis polynomial.
Such a matrix element of the normal product is a finite Laurent polynomial.
The factors $K_s(z_j/z_i)$ have convergent geometric expansions on
equal-radius circles, since $|s z_j/z_i|<1$.  Ordered radii, chosen
close enough to each other, give the same constant term and may be
deformed to equal radii without crossing a pole.  These observations
justify the normal ordering and the permutation average as actual
coefficient calculations.  The resulting finite-degree matrix entries
are rational in $s$, so \eqref{eq:H-power} then continues
coefficientwise to $s=t^{-1}$.  We do not identify a contour integral
at $|s|>1$ with its original small-$s$ expansion.

\subsection{An elementary trace calculation}

The following calculation extends the free-field trace formula of
\cite[Proposition~2.3]{Koshida2021} to independent weights $g_n$
for the power-sum modes. We include its elementary binomial derivation
for completeness.

Fix $n\geq1$ and first calculate in the single power-sum variable
$p_n$. The operator \eqref{eq:T-def} acts on its powers by
$\mathcal T(p_n^m)=g_n^m p_n^m$ for $m\geq0$.
Let $A_n,B_n$ be commuting parameters. The diagonal coefficient of
$e^{A_np_n}e^{B_n\partial_{p_n}}$ acting on $p_n^m$ is
\begin{equation}\label{eq:one-trace-diagonal}
 \sum_{j=0}^m\frac{A_n^jB_n^j}{(j!)^2}\frac{m!}{(m-j)!}
 =\sum_{j=0}^m\frac{(A_nB_n)^j}{j!}\binom mj.
\end{equation}
Indeed, to return to the same monomial $p_n^m$, the number of
differentiations and multiplications must be equal. Here
$\binom mj=m!/[j!(m-j)!]$ is the binomial coefficient.
Differentiating the geometric series $\sum_{m\geq0}g_n^m=(1-g_n)^{-1}$
$j$ times gives
\begin{equation}\label{eq:binomial-geometric}
 \sum_{m\geq j}\binom mj g_n^m=\frac{g_n^j}{(1-g_n)^{j+1}}.
\end{equation}
Multiply \eqref{eq:one-trace-diagonal} by $g_n^m$ and sum over $m$.
Using \eqref{eq:binomial-geometric}, we find
\begin{equation}\label{eq:one-trace}
 \operatorname{Tr}_{\mathbb C[p_n]}
 \left(\mathcal T e^{A_np_n}e^{B_n\partial_{p_n}}\right)
 =\frac1{1-g_n}\exp\left(\frac{A_nB_ng_n}{1-g_n}\right).
\end{equation}
Here $\mathbb C[p_n]$ is the polynomial ring in the single variable $p_n$.
This is a sum of explicitly computed diagonal entries; no cyclic
rearrangement of an infinite-dimensional trace is involved.

The variables $p_n$ are independent, so their diagonal sums multiply.
Apply \eqref{eq:one-trace} to \eqref{eq:normal-product} with
\begin{equation}
 A_n=\frac{1-s^n}{n}p_n(z),\qquad
 B_n=-p_n(z^{-1}).
\end{equation}
After division by $R=\prod_n(1-g_n)^{-1}$, the result is
\begin{equation}\label{eq:normal-trace}
 \frac{\operatorname{Tr}_{\Lambda}
 \bigl(\mathcal T:\eta(z_1)\cdots\eta(z_k):\bigr)}{R}
 =\exp\left[-\sum_{n\geq1}
 \frac{(1-s^n)g_n}{n(1-g_n)}p_n(z)p_n(z^{-1})\right].
\end{equation}
For an operator that is not degree-preserving, the trace in this
formula means the sum of its diagonal degree blocks.  It is still
well-defined formally: a diagonal monomial of weighted degree $d$
carries weight $d$ in the $g_n$.  Thus finitely many diagonal entries
contribute at every specified weight.

Combining \eqref{eq:C-as-trace}, \eqref{eq:H-power}, and
\eqref{eq:normal-trace} gives
\begin{align}\label{eq:C-constant-term}
 C_k={}&\frac1{k!(1-s)^k}\operatorname{CT}_z\left\{
 \Delta_s(z)
 \exp\left[-\sum_{n\geq1}
 \frac{(1-s^n)g_n}{n(1-g_n)}p_n(z)p_n(z^{-1})\right]
 \right\}.
\end{align}
The same calculation that gave \eqref{eq:hl-measure} shows that
$\Delta_s(z)/[k!(1-s)^k]$ is the unitary measure together with the
exponential generated by the pure-boson sequence $s^n$.
Consequently \eqref{eq:C-constant-term} is precisely
\begin{equation}\label{eq:dual-couplings}
 C_k=Z_k[\widehat f],\qquad
 \widehat f_n=s^n-\frac{(1-s^n)g_n}{1-g_n}
             =1-\frac{1-s^n}{1-g_n}.
\end{equation}
For $k=0$ the formula reads $C_0=Z_0=1$.

We have therefore proved the general transformation proposed by
Gaiotto and Lee \cite[Eqs.~(1.6) and~(1.11)]{Gaiotto:2021xce}:
\begin{equation}\label{eq:general-gge}
 \boxed{\displaystyle
 \frac{Z_N[f]}{Z_\infty[f]}
 =\sum_{k=0}^{\infty}t^{kN}Z_k[\widehat f],\qquad
 (1-f_n)(1-\widehat f_n)=(1-t^n)(1-t^{-n}).}
\end{equation}
The last relation follows directly from \eqref{eq:g-def} and
\eqref{eq:dual-couplings}.  It identifies every coefficient of the
rank expansion, rather than assuming that the coefficients have the
form of an index.

\subsection{Application to the full index and analytic continuation}

For the sequence \eqref{eq:full-letter}, the variables in
\eqref{eq:g-def} are
\begin{equation}\label{eq:physical-g}
 g_n=1-\frac{(1-u^n)(1-v^n)}{(1-p^n)(1-q^n)}.
\end{equation}
Substituting into \eqref{eq:general-large-rank} proves
\eqref{eq:full-large-rank}.  Substituting into
\eqref{eq:dual-couplings} gives
\begin{equation}\label{eq:physical-dual}
 \widehat f_n
 =1-\frac{(1-t^{-n})(1-p^n)(1-q^n)}{(1-u^n)(1-v^n)}.
\end{equation}
This is exactly \eqref{eq:full-letter} with the replacement
\begin{equation}
 (t,u,v;p,q)\longmapsto(t^{-1},q,p;v,u).
\end{equation}
The constraint is respected, because $pq=tuv$ implies
$t^{-1}qp=vu$.  Therefore
$Z_k[\widehat f]=I_k(t^{-1},q,p;v,u)$, with the continuation already
specified, and \eqref{eq:general-gge} proves \eqref{eq:full-gge}.

We make the logical status of the continuation and summation explicit.
At fixed weight in the independent $g_n$, all changes of polynomial
basis and all traces involve only finite-dimensional spaces.  The
coefficients obtained from them are rational functions of $t$ (or of
$s=t^{-1}$).  The Hall--Littlewood matrix integral is initially
evaluated for $|t|<1$, whereas the dual constant term in
\eqref{eq:C-constant-term} is initially evaluated for $|s|<1$.
Their rational coefficients can be continued between these domains;
there is no assertion that both original contour integrals converge
on the same domain.  Equation \eqref{eq:rank-coefficients}, followed by
\eqref{eq:dual-couplings}, identifies the continued dual coefficients
with the Taylor coefficients of $\Phi(\xi)$.

For $0<|t|<1$, each auxiliary coefficient of $\Phi(\xi)$ is entire,
as proved after \eqref{eq:interpolation}.  Its Taylor series therefore
converges at $\xi=t^N$, justifying the infinite sum over $k$.
Upon the physical substitution \eqref{eq:physical-scaling},
\eqref{eq:physical-g} satisfies $g_n=O(\varepsilon^n)$.
At order $\varepsilon^d$, only finitely many monomials in the $g_n$ of
weight at most $d$ contribute.  The established coefficientwise
identities and convergence thus persist after the physical
substitution.  This proves the full-index single-sum expansion to every
order in the simultaneous small-fugacity expansion.

As a simple normalization check, setting every $g_n=0$ gives
$Z_N=1/(t;t)_N$ and $\widehat f_n=t^{-n}$.  The result becomes
\begin{equation}
 \frac{(t;t)_\infty}{(t;t)_N}
 =\sum_{k=0}^{\infty}\frac{t^{kN}}{(t^{-1};t^{-1})_k}
 =\sum_{k=0}^{\infty}
 \frac{(-1)^kt^{kN+k(k+1)/2}}{(t;t)_k},
\end{equation}
which is exactly the Euler expansion of $(t^{N+1};t)_\infty$.

\section{The double-sum expansion of the Schur index}\label{sec:schur}

The Schur index admits a giant graviton expansion with two independent
wrapping numbers \cite{Arai:2020qaj,Beccaria:2023zjw,Beccaria:2024szi}.  This section proves that expansion at arbitrary
rank.  Its proof uses the exact free-fermion representation of the Schur
index, rather than a specialization of the single-sum expansion in
Section~\ref{sec:proof}.  The distinction matters because the two
expansions use different orders of expansion in the fugacities.  We
will specify the continuation and prove that only finitely many wrapping
sectors contribute at each fixed order in the physical radial variable.

\subsection{The Schur index and the two wrapping numbers}

The Schur specialization \cite{Gadde:2011uv} sets $v=p$, so the balancing relation becomes
\begin{equation}\label{eq:schur-balance}
 q=tu.
\end{equation}
We keep the same fugacities $t,u$ and use the two-argument notation
\begin{equation}\label{eq:schur-full-specialization}
 I_N(t,u):=I_N(t,u,p;\,p,tu).
\end{equation}
The factors involving $p$ cancel in \eqref{eq:full-letter}, so this
specialized index is independent of $p$. Its matrix integral is
\begin{equation}\label{eq:schur-haar}
 I_N(t,u)=\int_{U(N)}\dd U\,
 \exp\left[
 \sum_{n=1}^{\infty}\frac{f_n}{n}
 \Tr(U^n)\Tr(U^{-n})\right],
 \qquad I_0(t,u)=1,
\end{equation}
where the same sequence $f_n$ now specializes to
\begin{equation}\label{eq:schur-letter}
 f_n=\frac{t^n+u^n-2q^n}{1-q^n}
 =1-\frac{(1-t^n)(1-u^n)}{1-q^n},\qquad q=tu.
\end{equation}
The integral is initially understood for $|t|,|u|<1$, or as a power
series at $t=u=0$.  It is unchanged by interchanging $t$ and $u$.
Its large-rank limit is
\begin{equation}\label{eq:schur-large-rank}
 I_\infty(t,u)=\prod_{k=1}^{\infty}
 \frac{1-q^k}{(1-t^k)(1-u^k)}
 =\frac{(q;q)_\infty}{(t;t)_\infty(u;u)_\infty}.
\end{equation}
The infinite products use the notation of
\eqref{eq:pochhammer-def}.

The result to be proved is
\begin{equation}\label{eq:schur-double-sum}
 \boxed{\displaystyle
 \frac{I_N(t,u)}{I_\infty(t,u)}
 =\sum_{a,b\ge0}t^{aN}u^{bN}q^{ab}
 I_a(t^{-1},q)\,I_b(q,u^{-1}) ,\qquad N\ge0.}
\end{equation}
The nonnegative integers $a,b$ count the two species of wrapped giant
gravitons.  The factors $t^{aN}$ and $u^{bN}$ record their classical
charges, and $q^{ab}$ couples the two wrapping numbers.  The two
finite-rank indices on the right are understood by meromorphic
continuation in their arguments, with respective nomes $u$ and $t$.
We work with $0<|t|,|u|<1$, away from individual poles; substituting
inverse fugacities into their original Taylor series does not define
these factors.  The restriction to nomes of modulus less than one is
substantive: already at rank one, poles accumulate at every point of the nome unit circle for a
fixed generic fugacity, obstructing meromorphic continuation across it.
The continuation used here is constructed below.

To state the expansion precisely, put
\begin{equation}\label{eq:schur-radial-scaling}
 t=\varepsilon t_0,\qquad u=\varepsilon u_0,
 \qquad q=\varepsilon^2t_0u_0,
\end{equation}
where $t_0,u_0$ are fixed nonzero generic parameters and $\varepsilon$ is a
bookkeeping variable.  Equality means equality of every coefficient of
$\varepsilon$, with coefficients rational in $t_0,u_0$.
Generic parameters avoid the poles of intermediate expressions, such as
$t_0^k=u_0^k$.  A special flavor value is taken after the finite sum
contributing to a specified coefficient has been formed.  We will prove
that the term with wrapping numbers $(a,b)$ starts no earlier than
\begin{equation}\label{eq:schur-term-order}
 \varepsilon^{(a+b)^2+N(a+b)}.
\end{equation}
Thus only finitely many pairs $(a,b)$ contribute to any given radial
order.  This is the precise sense in which the double sum is defined.

Equation \eqref{eq:schur-double-sum} is Eq.~(1.4) of
\cite{Beccaria:2024szi}, written in our fugacities and with the two
wrapping numbers summed independently.

\subsection{An exact generating function from the Fermi gas}

We use the multiplicative theta function
\begin{equation}\label{eq:schur-theta-definition}
 \theta(w;r)=(w;r)_\infty(r/w;r)_\infty,
 \qquad w\ne0,\quad 0<|r|<1,
\end{equation}
so that its normalization contains no factor $(r;r)_\infty$.
For later convenience we exclude $r=0$.
The variable $r$ in this definition denotes a general nome, meaning
the common ratio in the infinite products.  In applications below it
will be $q$, $t$, or $u$.

The free-fermion representation of the Schur index was developed in
\cite{Bourdier:2015wda,Hatsuda:2022xdv}; a useful normalization of its
generating function also appears in Appendix~B of
\cite{Gaiotto:2021xce}.  We give the exact identity in our conventions
and explain its reduction to the matrix integral, since its prefactors
will be needed in the coefficient calculation.

For auxiliary variables $\zeta,w$, define
\begin{equation}\label{eq:schur-F-definition}
 \mathcal F_{t,u}(\zeta,w)
 =\theta(w;q)\prod_{k\in\mathbb Z}
 \left(1+\frac{\zeta t^k}{1-wq^k}\right).
\end{equation}
The product over $k$ is the grand partition function of auxiliary free
fermions: the two terms in each factor represent an empty or singly
occupied mode, and $\zeta$ is the particle-number fugacity. The theta
prefactor fixes the normalization, and $w$ is an auxiliary parameter
that will cancel from the fixed-rank index.
For $|t|,|u|<1$, the factors approach $1$ geometrically in both
directions of $k$, provided $w$ avoids the displayed poles.  Hence the
product is convergent.  The coefficient identity is
\begin{equation}\label{eq:schur-F-coefficient}
 \mathcal F_{t,u}(\zeta,w)
 =\sum_{n=0}^{\infty}\zeta^n t^{n(n-1)/2}
 \theta(wt^n;q)I_n(t,u).
\end{equation}
Thus the gauge rank $n$ is encoded by the fermion particle number, with
the displayed prefactors relating the index to the fixed-particle sector.

Here is a derivation.  Diagonalizing $U$ in
\eqref{eq:schur-haar}, expanding the exponential into infinite
products, and pairing factors with arguments $z_i/z_j$ and $z_j/z_i$
gives
\begin{equation}\label{eq:schur-theta-integral}
 I_n(t,u)=\frac{(q;q)_\infty^{2n}}{n!\,\theta(t;q)^n}
 \oint\prod_{i=1}^n\frac{\dd z_i}{2\pi\mathrm{i}z_i}
 \prod_{i\ne j}
 \frac{\theta(z_i/z_j;q)}{\theta(tz_i/z_j;q)}.
\end{equation}
The factor $1/n!$ is the usual permutation factor in the Haar
integration formula.  Each $z_i$ is integrated around the unit circle.
Introduce the Fourier kernel
\begin{equation}\label{eq:schur-fourier-kernel}
 K(z)=\sum_{k\in\mathbb Z}\frac{t^kz^k}{1-wq^k}
 =\frac{(q;q)_\infty^2\theta(wtz;q)}
 {\theta(w;q)\theta(tz;q)}.
\end{equation}
The second equality is the classical bilateral theta summation.
Its Fourier coefficients, namely the weights multiplying $z^k$, are
$\lambda_k=t^k/(1-wq^k)$.  The multiplicative Cauchy determinant for
theta functions then gives
\begin{align}
 \det_{1\le i,j\le n}K(z_i/z_j)
 &=\frac{t^{n(n-1)/2}\theta(wt^n;q)}{\theta(w;q)}
 \frac{(q;q)_\infty^{2n}}{\theta(t;q)^n}\notag\\
 &\quad\times
 \prod_{i\ne j}
 \frac{\theta(z_i/z_j;q)}{\theta(tz_i/z_j;q)}.
 \label{eq:schur-frobenius}
\end{align}
To obtain the generating product, first truncate the Fourier series of
$K$ to finitely many integers $k$.  The finite Cauchy--Binet formula
expands the determinant as a sum over subsets of $n$ distinct modes:
\begin{equation}\label{eq:schur-cauchy-binet}
 \det[K(z_i/z_j)]
 =\sum_{k_1<\cdots<k_n}
 \lambda_{k_1}\cdots\lambda_{k_n}
 \det[z_i^{k_a}]_{i,a}\det[z_j^{-k_a}]_{a,j},
\end{equation}
where the sum is over the retained modes. The two determinants are
Slater determinants of Fourier modes. Repeated modes give identical
columns in the first determinant and identical rows in the second,
so their contributions vanish, implementing the Pauli exclusion principle.
Integrating each $z_i$ around the unit circle and expanding the two
finite determinants shows that the integral of their product is $n!$:
the identity $\oint z^m\,\dd z/(2\pi\mathrm{i}z)=\delta_{m,0}$, where
$\delta_{m,0}$ is $1$ for $m=0$ and $0$ otherwise, pairs only equal
permutations of the distinct exponents.  Therefore the integral of the
kernel determinant divided by $n!$ equals the $n$th elementary symmetric
function of the $\lambda_k$.
For $|q|<|t|<1$, the Fourier coefficients are absolutely summable and
the Fourier series converges uniformly on the unit circle, so the
truncation can be removed.  The resulting sum
$\sum_{k_1<\cdots<k_n}\lambda_{k_1}\cdots\lambda_{k_n}$ is the coefficient
of $\zeta^n$ in $\prod_{k\in\mathbb Z}(1+\zeta\lambda_k)$.
Comparing \eqref{eq:schur-frobenius} with
\eqref{eq:schur-theta-integral} proves
\eqref{eq:schur-F-coefficient}. Appendix~\ref{app:schur-theta-proofs}
recalls Jacobi's triple product and proves the bilateral summation and
the determinant identity by elementary complex-analysis arguments.

There is also a useful product formula without denominators.
Cancel the factors in $\theta(w;q)$ separately against the $k\ge0$
and $k<0$ factors in \eqref{eq:schur-F-definition}.  The result is
\begin{equation}\label{eq:schur-F-trinomial}
 \mathcal F_{t,u}(\zeta,w)
 =\prod_{n\ge0}(1+\zeta t^n-wq^n)
 \prod_{m\ge1}\left(1-\frac{q^m+\zeta u^m}{w}\right).
\end{equation}
This elementary rewriting is the starting point of the factorization.

\subsection{A coefficientwise proof of the factorization}
\label{sec:schur-formal-factorization}

Let $A,B$ be independent auxiliary variables.  We will prove
\begin{equation}\label{eq:schur-factorization}
 \mathcal F_{t,u}(A,-AB)
 =A\,\mathcal F_{t^{-1},q}(A^{-1},-B)
       \mathcal F_{u^{-1},q}(B,-A^{-1}).
\end{equation}
On the right, each fixed-rank coefficient is understood by meromorphic
continuation of \eqref{eq:schur-F-coefficient}.  We prove the identity
coefficientwise under the radial scaling \eqref{eq:schur-radial-scaling};
the uncontinued products have no common convergence domain.
Appendix~\ref{app:schur-continuation} verifies that the formal product
prescription below computes the radial expansions of these continued
coefficients.

Introduce four factors:
\begin{align}
 P_1(A,B)&=\prod_{n\ge0}(1+A^{-1}t^{-n}+Bu^n),
\notag\\
 P_2(A,B)&=\prod_{n\ge0}(1+A^{-1}t^n+Bu^{-n}),
 \label{eq:schur-P-products}\\
 C_+(A,B)&=\prod_{m\ge1}(1+At^m+ABq^m),
\notag\\
 C_-(A,B)&=\prod_{m\ge1}
 (1+B^{-1}u^m+A^{-1}B^{-1}q^m).
 \label{eq:schur-XY-products}
\end{align}
The tail products $C_+,C_-$ are well defined as formal series in
$\varepsilon$, since the $m$th factor differs from $1$ only in degrees
at least $m$.  We define $P_1,P_2$ coefficientwise in nonnegative
powers of $A^{-1}$ and $B$ through their logarithms.  For $i,j\ge0$
and $i+j>0$, the coefficient of $A^{-i}B^j$ in $\log P_1$ is
\begin{equation}\label{eq:schur-log-P1}
 \frac{(-1)^{i+j+1}}{i+j}\binom{i+j}{i}\frac1{1-t^{-i}u^j},
\end{equation}
and in $\log P_2$ it is
\begin{equation}\label{eq:schur-log-P2}
 \frac{(-1)^{i+j+1}}{i+j}\binom{i+j}{i}\frac1{1-t^iu^{-j}}.
\end{equation}
These rational expressions follow by expanding the logarithm and
summing geometric series where the products converge; elsewhere they
define the coefficientwise continuation.

To interpret these coefficients as radial series, we prescribe their
Taylor expansion at $\varepsilon=0$ after substituting
$t=\varepsilon t_0$, $u=\varepsilon u_0$.
In \eqref{eq:schur-log-P1}, when $j<i$ we first rewrite
$(1-t^{-i}u^j)^{-1}=-t^iu^{-j}/(1-t^iu^{-j})$ and then expand
geometrically in $t^iu^{-j}$, whose radial order is $i-j>0$.
When $j>i$, we expand in $t^{-i}u^j$; when $i=j$, we retain the
$\varepsilon$-independent rational factor, which is finite by genericity.
The same rule applies to \eqref{eq:schur-log-P2}, with the inequalities
reversed. These rewritings exhibit the rational coefficients as regular
at $\varepsilon=0$, so their prescribed Taylor series have only
nonnegative radial powers. Expanding geometrically in a quantity of
negative radial order would not give a formal power series in
$\varepsilon$.
This expansion prescription is part of the formal definition of the
continued products $P_1,P_2$ in the radial expansion. Formal
exponentiation is well defined because the logarithms have zero
constant term in $A^{-1},B$, so each coefficient in these variables
requires only finitely many operations.

The elementary identity
\begin{equation}\label{eq:schur-geometric-cancellation}
 \frac1{1-r}+\frac1{1-r^{-1}}=1
\end{equation}
therefore gives, with this prescription, the formal power-series identity
\begin{equation}\label{eq:schur-P-cancellation}
 \log P_1+\log P_2=\log(1+A^{-1}+B),
 \qquad P_1P_2=1+A^{-1}+B.
\end{equation}

With this prescription, each radial coefficient of $P_1,P_2$ is an
ordinary formal power series in $A^{-1},B$. This allows regrouping
with the tails as follows.

At any fixed radial order, $C_+C_-$ contributes only finitely many
Laurent monomials in $A,B$.  For each such monomial and a prescribed
coefficient $[A^\alpha B^\beta]$, the total nonnegative powers of
$A^{-1}$ and $B$ supplied by $P_1,P_2$ are fixed and have only finitely
many splittings between the two factors.  Hence every
$[\varepsilon^D A^\alpha B^\beta]$ coefficient uses only finitely many
terms, and the four factors can be regrouped formally.

Finally, \eqref{eq:schur-F-trinomial} gives, with the continued
coefficient interpretation just specified,
\begin{align}
 \mathcal F_{t,u}(A,-AB)&=(1+A+AB)C_+C_- ,\notag\\
 \mathcal F_{t^{-1},q}(A^{-1},-B)&=P_1C_- ,\notag\\
 \mathcal F_{u^{-1},q}(B,-A^{-1})&=P_2C_+ .
 \label{eq:schur-four-factor-identification}
\end{align}
Using \eqref{eq:schur-P-cancellation},
\[
 A(P_1C_-)(P_2C_+)
 =A(1+A^{-1}+B)C_+C_-
 =(1+A+AB)C_+C_-.
\]
This proves \eqref{eq:schur-factorization} coefficient by coefficient.

\subsection{Extracting the two-species expansion}

The remaining calculation is an explicit coefficient extraction.
With our theta normalization, the Jacobi triple product reads
\begin{equation}\label{eq:schur-triple-product}
 \theta(-z;r)=\frac1{(r;r)_\infty}
 \sum_{k\in\mathbb Z}r^{k(k-1)/2}z^k.
\end{equation}
We take $[A^NB^0]$ on both sides of
\eqref{eq:schur-factorization}.  On the left,
\eqref{eq:schur-F-coefficient} gives
\[
 \mathcal F_{t,u}(A,-AB)
 =\sum_{n\ge0}A^n t^{n(n-1)/2}
 \theta(-ABt^n;q)I_n(t,u).
\]
In \eqref{eq:schur-triple-product}, the power of $B$ is its theta
summation index.  Thus $B^0$ forces that index to be zero, and $A^N$
then forces $n=N$.  The left coefficient is
\begin{equation}\label{eq:schur-left-coefficient}
 \frac{t^{N(N-1)/2}}{(q;q)_\infty}I_N(t,u).
\end{equation}

On the right, the expansions in the two particle numbers are
\begin{align}
 \mathcal F_{t^{-1},q}(A^{-1},-B)
 &=\sum_{a\ge0}A^{-a}t^{-a(a-1)/2}
 \theta(-Bt^{-a};u)I_a(t^{-1},q),\notag\\
 \mathcal F_{u^{-1},q}(B,-A^{-1})
 &=\sum_{b\ge0}B^b u^{-b(b-1)/2}
 \theta(-A^{-1}u^{-b};t)I_b(u^{-1},q).
 \label{eq:schur-two-fermion-expansions}
\end{align}
Let $\nu_1,\nu_2$ denote the two integer theta summation indices.
Including the prefactor $A$ in
\eqref{eq:schur-factorization}, the total powers of $A,B$ are
$1-a-\nu_2$ and $b+\nu_1$.  Therefore the requested coefficient forces
\begin{equation}\label{eq:schur-theta-indices}
 \nu_1=-b,\qquad \nu_2=1-a-N.
\end{equation}
For this pair of integers the exponent of $t$ is
\begin{align}
 -\frac{a(a-1)}2+ab+\frac{(a+N-1)(a+N)}2
 =\frac{N(N-1)}2+aN+ab,
 \label{eq:schur-x-exponent}
\end{align}
and the exponent of $u$ is
\begin{equation}\label{eq:schur-y-exponent}
 -\frac{b(b-1)}2+\frac{b(b+1)}2-b(1-a-N)
 =bN+ab.
\end{equation}
The right coefficient is consequently
\begin{equation}\label{eq:schur-right-coefficient}
 \frac{t^{N(N-1)/2}}{(t;t)_\infty(u;u)_\infty}
 \sum_{a,b\ge0}t^{aN}u^{bN}q^{ab}
 I_a(t^{-1},q)I_b(u^{-1},q).
\end{equation}
Comparing \eqref{eq:schur-left-coefficient} and
\eqref{eq:schur-right-coefficient}, canceling the common factor,
and using the symmetry $I_b(u^{-1},q)=I_b(q,u^{-1})$ proves
\eqref{eq:schur-double-sum}.
The two wrapping numbers are therefore the fixed-particle labels $a,b$
of the continued fermion factors. Their mixed weight $q^{ab}$ comes
from the theta modes selected by the common Laurent projection.

The coefficientwise finiteness established in
Section~\ref{sec:schur-formal-factorization} also justifies grouping by
particle number.  Each contributing product monomial has a definite
$A^{-1}$-degree $a$ in $P_1C_-$ and $B$-degree $b$ in $P_2C_+$.
Thus only finitely many such groups contribute to each prescribed
radial and Laurent coefficient.

\subsection{Radial order, specializations, and the single-sum expansion}

We finish by proving the order estimate
\eqref{eq:schur-term-order}, which also makes the physical meaning of
the answer precise.  Say that a series has radial order at least $d$
if its expansion in $\varepsilon$ contains no power below
$\varepsilon^d$.  Every coefficient of $P_1C_-$ has nonnegative
radial order.  Take $[A^{-a}B^a]$ in the first equation of
\eqref{eq:schur-two-fermion-expansions}.
By \eqref{eq:schur-triple-product}, this coefficient is
\begin{equation}\label{eq:schur-brane-order-coefficient}
 \frac{t^{-a(a-1)/2-a^2}u^{a(a-1)/2}}
 {(u;u)_\infty}\,I_a(t^{-1},q).
\end{equation}
The prefactor has radial order $-a^2$, and its remaining factor has a
nonzero constant term.  Hence
\begin{equation}\label{eq:schur-brane-order}
 I_a(t^{-1},q)=O(\varepsilon^{a^2}),\qquad
 I_b(q,u^{-1})=O(\varepsilon^{b^2}).
\end{equation}
Multiplication by $t^{aN}u^{bN}q^{ab}$ gives exactly
\eqref{eq:schur-term-order}.  This bound was derived from the
continued generating functions and is not an assumption about the
giant graviton expansion.

The coefficient extraction also gives a practical method for computing
the radial expansions of the continued brane indices. Direct expansion
of the transformed matrix integral \eqref{eq:schur-theta-integral}
requires a prescription for continuing its contours. Equations
\eqref{eq:schur-two-fermion-expansions} and
\eqref{eq:schur-brane-order-coefficient} recover the indices from
Laurent coefficients of $P_1C_-$ and $P_2C_+$, computed using the
rational coefficients of $P_1,P_2$ and the tail products $C_-,C_+$.
Each fixed radial coefficient requires only finitely many operations
and is a rational function of $t_0,u_0$.
Appendix~\ref{app:schur-continuation} identifies these series with the
meromorphic continuations of the original indices. This construction
requires no quasimodular representation. In \cite{Beccaria:2024szi},
exact theta-function and twisted-Weierstrass representations implement
the continuation, while quasimodular forms organize the near-unflavored
expansion.

Together with the bound \eqref{eq:schur-term-order}, this method permits
explicit finite-order checks of \eqref{eq:schur-double-sum} against an
independent expansion of the original matrix integral. At $t=u$, possible
poles in individual brane coefficients cancel in the finite sum at each
radial order, so this specialization is taken only after summation.

A simple check is the half-BPS specialization of the total index.
Setting $u=0$ in \eqref{eq:schur-haar} gives
$I_N(t,0)=1/(t;t)_N$ and $I_\infty(t,0)=1/(t;t)_\infty$.
The normalized answer is then
\begin{equation}\label{eq:schur-half-bps-check}
 \frac{I_N(t,0)}{I_\infty(t,0)}
 =(t^{N+1};t)_\infty
 =\sum_{a\ge0}\frac{(-1)^a t^{aN+a(a+1)/2}}{(t;t)_a}.
\end{equation}
The final equality is Euler's product expansion.  When checking the
double-sum formula in such a limit, the total continued expression
must be specialized; the proof does not require assigning independent
limits to divergent or singular intermediate factors.

Finally, the double sum proved here uses
$t=\varepsilon t_0,u=\varepsilon u_0$, so both physical Schur fugacities
become small together.  The Gaiotto--Lee single-sum expansion uses a
different coefficientwise completion, with one fugacity kept fixed
while the other small variables are expanded.  A formal identity in
that completion does not by itself justify exchanging the order of
expansion to obtain \eqref{eq:schur-double-sum}.  The factorization and
finite coefficient counting above provide the independent justification
for the physical two-species expansion.

\section{Conclusion}\label{sec:conclusion}

We have proved two giant graviton identities with different ranges of
application. Hall--Littlewood orthogonality and the length-measuring
operator establish the Gaiotto--Lee single-sum expansion for the
full $\mathcal N=4$ $U(N)$ index. An independent factorization of
the Schur free-fermion product establishes the symmetric double-sum
formula at every rank and wrapping number. In both arguments the
brane coefficients are obtained from the matrix index with a stated
analytic continuation. The justification of the sums is
coefficientwise: it does not assume uniform convergence in every
region of complex fugacity space.

The most immediate open problem is the full three-species expansion
\eqref{eq:intro-imamura}. Its resolution should identify a principle
that determines the worldvolume integration cycles and consistently
organizes interactions among three intersecting brane stacks at
arbitrary wrapping. Such a principle would clarify why a gauge-theory
index reorganizes into the degrees of freedom of coupled D3-branes.
Neither the single-sum theorem nor the special Schur factorization
provides this missing identification. Symmetric-function
representations of deformed indices and of the full elliptic index
\cite{Ren:2025tvx,Ren:2026dlj} offer useful exact information, but
connecting them to the physical three-node brane theory requires
additional ideas.

Applications of giant graviton expansions to black-hole microstates
\cite{Choi:2022ovw} also require control of large-charge limits.
Contributions with wrapping numbers that grow with $N$, and the complex-fugacity saddle points
relevant to fixed-charge degeneracies, need not be controlled by a
coefficientwise expansion around small fugacities. A useful next
step is to determine when the sum over branes can be combined with
these large-charge limits and how the resulting collective saddle
is represented in the bulk. This requires control of cancellations,
integration cycles, and backreaction in the full index. The Schur
limit gives an exact laboratory for some of these mechanisms, while
the more general supersymmetric black-hole problem involves the
additional charges retained by the full index. Existing modular
and high-temperature analyses \cite{ArabiArdehali:2023bpq} provide
useful constraints on the corresponding limits of the Schur index.

Another direction is to extend the symmetric-function approach to
M-theory. Hayashi, Nosaka and Okazaki obtained M2--M5 giant graviton
expansions for Coulomb and Higgs indices of M2-brane theories, relating
them to protected M5-brane spectra and twisted indices of
six-dimensional $(2,0)$ theories \cite{Hayashi:2024aaf}.
Hwang, Lei and Tang studied finite-$N$ M2-brane indices using the
factorization formula for the three-dimensional ADHM quiver theory
and explored their implications for black-hole microstates
\cite{Hwang:2025hfs}. These works provide concrete examples and index
data on which to build an M-theory generalization. A central question
is whether an appropriate symmetric-function description can both
derive the brane expansion and identify its coefficients with the
protected spectra of M2- and M5-branes.

Finally, the modular properties of the Schur index suggest a
particularly concrete bridge between branes and the protected
two-dimensional algebra. It would be valuable to determine how
modular transformations reorganize wrapping sectors, and whether
the continued worldvolume indices have a direct interpretation in
terms of modules or defects of that algebra. Related character
identifications already exist in a different setting: Dorey and Roehl
related giant graviton coefficients in Higgs- and Coulomb-branch limits
of three-dimensional $\mathcal N=4$ theories to characters of
parafermionic $W$-algebras \cite{Dorey:2025qht}. The question here is
to identify the corresponding four-dimensional Schur sectors and their
modular transformations, including the interaction between the two
wrapping numbers. The same question can
be asked with line operators and boundaries, where exact indices
and giant graviton constructions are already available
\cite{Hatsuda:2023iwi,Hatsuda:2024uwt}. A physical identification of the relevant
states and sectors could turn modular anomaly equations from
efficient computational tools into constraints on the bulk brane
description. These questions require physical information beyond
formal rearrangements of an index and are natural directions for
further work.

\vspace{0.2in} {\leftline {\bf Acknowledgments}}
\nopagebreak
The derivations and proofs in this paper were developed with assistance from GPT-6 Astra. The text was written by the author with assistance from AI tools. The author is responsible for checking the arguments and for the final manuscript. We thank  Gao-fu Ren, Xin Wang for helpful discussions. This work is  supported by the National Natural Science Foundation of China Grants No. 12325502 and No. 12247103.  

\appendix

\section{Elementary theta-function identities}
\label{app:schur-theta-proofs}
This appendix collects the theta identities used in the free-fermion
generating formula and the final coefficient extraction.  We recall
Jacobi's triple product and prove the bilateral summation and elliptic
Cauchy determinant using Laurent series, the residue theorem, and
elementary properties of elliptic functions. We use the theta function
\eqref{eq:schur-theta-definition}, with nome $r$ satisfying $0<|r|<1$
and nonzero argument $z$. The notation
$\mathbb C^*=\mathbb C\setminus\{0\}$ denotes the punctured complex
plane. A prime on $\theta$ differentiates its first argument.
A holomorphic function is one that is complex analytic at every point
of its stated domain. The product definition immediately gives
\begin{equation}\label{eq:theta-elementary-rules}
 \theta(rz;r)=-z^{-1}\theta(z;r),\qquad
 \theta(z^{-1};r)=-z^{-1}\theta(z;r),\qquad
 \theta'(1;r)=-(r;r)_\infty^2.
\end{equation}
The zeros are the points $z=r^m$, $m\in\mathbb Z$, and are simple.
The derivative identity follows by separating the factor $(1-z)$ from
the defining product and then setting $z=1$.

We also recall the standard Jacobi triple product, which in our
normalization reads
\begin{equation}\label{eq:jacobi-triple-product-proved}
 \theta(-z;r)=\frac1{(r;r)_\infty}
 \sum_{n\in\mathbb Z}r^{n(n-1)/2}z^n.
\end{equation}
The series converges absolutely for $0<|r|<1$ and $z\ne0$.

\subsection{The bilateral Fourier identity}
Write $r^{\mathbb Z}=\{r^m:m\in\mathbb Z\}$.
For $|r|<|z|<1$ and $w\in\mathbb C^*\setminus r^{\mathbb Z}$, one has
\begin{equation}\label{eq:kronecker-proved}
 \frac{(r;r)_\infty^2\theta(wz;r)}{\theta(w;r)\theta(z;r)}
 =\sum_{n\in\mathbb Z}\frac{z^n}{1-wr^n}.
\end{equation}
Here and below the arguments can subsequently be continued
meromorphically, with the nome restricted to $0<|r|<1$.

To prove this, denote the left side by $H(z)$ and let $c_n$ be its
Laurent coefficient on the annulus $|r|<|z|<1$.  Choose
$|r|<\rho<1$.  The first identity in \eqref{eq:theta-elementary-rules}
gives $H(rz)=w^{-1}H(z)$.  Consequently
\[
 \frac1{2\pi\mathrm{i}}\oint_{|z|=|r|\rho}H(z)z^{-n}\frac{\dd z}{z}
 =w^{-1}r^{-n}c_n.
\]
Only the pole $z=r$ lies between this contour and $|z|=\rho$.
The residue of $H(z)\,\dd z/z$ at $z=1$ is $-1$ by
$\theta'(1;r)=-(r;r)_\infty^2$.  The same functional equation shows that the
residue at $z=r$ is $-w^{-1}$.  The residue theorem therefore gives
\[
 c_n-w^{-1}r^{-n}c_n=-w^{-1}r^{-n},
 \qquad c_n=\frac1{1-wr^n}.
\]
This proves the claimed Laurent expansion.  Its normal convergence
on compact subannuli also follows directly: the positive-$n$ tail is
bounded by a geometric series in $|z|$, and the negative-$n$ tail by
one in $|r/z|$.

\subsection{The elliptic Cauchy determinant}
Let $n$ be a positive integer, and let $w,x_i,y_j$
($1\le i,j\le n$) be generic nonzero complex parameters.
The following finite determinant identity holds:
\begin{equation}\label{eq:frobenius-proved}
 \det_{1\leq i,j\leq n}
 \left[\frac{\theta(w x_i y_j;r)}
 {\theta(w;r)\theta(x_i y_j;r)}\right]
 =\frac{\theta(w\prod_{i=1}^n x_i y_i;r)}{\theta(w;r)}
 \frac{\displaystyle\prod_{i<j}x_jy_j\,
       \theta(x_i/x_j;r)\theta(y_i/y_j;r)}
 {\displaystyle\prod_{i,j}\theta(x_i y_j;r)}.
\end{equation}
It suffices to work with generic parameters, so that the zero and pole
orbits below are distinct; the remaining cases follow by meromorphic
continuation.

We prove \eqref{eq:frobenius-proved} by induction on $n$, the case
$n=1$ being immediate.  Denote its two sides by $L(z)$ and $R(z)$,
with $z=x_n$.  By \eqref{eq:theta-elementary-rules}, both satisfy
\begin{equation}\label{eq:frobenius-quasiperiod}
 F(rz)=w^{-1}F(z),\qquad F=L,R.
\end{equation}
Thus $Q(z)=L(z)/R(z)$ is $r$-periodic and defines a meromorphic function
on the compact torus $\mathbb C^*/r^{\mathbb Z}$.

At $z=r^m/y_j$, the function $R$ has a simple pole, while $L$ has at
most a simple pole because only one entry of its last row is singular.
Hence $Q$ is regular there.  At $z=r^m x_i$ with $i<n$, the function
$R$ has a simple zero, and the last determinant row is $w^{-m}$ times
row $i$.  Thus $L$ vanishes at least to first order, again making $Q$
regular.  These multiplicity bounds suffice; equality of multiplicities
need not be assumed.

The remaining numerator factor in $R$, namely $\theta(az;r)$ with
$a=w\prod_{i<n}x_i\prod_{j=1}^n y_j$, has one simple zero on the torus.
Thus $Q$ has at most one remaining simple pole.  The residue theorem applied to
$Q(z)\,\dd z/z$ excludes a single simple pole.  Therefore $Q$ is
holomorphic on the compact torus and is constant.

To fix the constant, compare residues at $z=1/y_n$.  The determinant
residue is its $(n-1)$-dimensional minor divided by $y_n\theta'(1;r)$.
On the product side, the new cross factors cancel at $x_ny_n=1$,
leaving the size-$(n-1)$ product divided by the same factor.
The induction hypothesis makes these residues equal and nonzero for
generic parameters, so $Q=1$.

With the Schur fugacities $t,u$ and $q=tu$ of Section~\ref{sec:schur},
set $r=q$, $x_i=tz_i$, and $y_j=z_j^{-1}$ in
\eqref{eq:frobenius-proved}, and multiply every matrix entry by
$(q;q)_\infty^2$.
The resulting kernel is
\[
 K(z)=\frac{(q;q)_\infty^2\theta(wtz;q)}{\theta(w;q)\theta(tz;q)}
 =\sum_{k\in\mathbb Z}\frac{t^kz^k}{1-wq^k},
\]
where the Fourier series converges for $|q|<|tz|<1$; elsewhere the
theta-function expression defines its meromorphic continuation.
The determinant identity becomes
\begin{equation}\label{eq:specialized-cauchy-proved}
 \det[K(z_i/z_j)]
 =t^{n(n-1)/2}\frac{\theta(wt^n;q)}{\theta(w;q)}
 \frac{(q;q)_\infty^{2n}}{\theta(t;q)^n}
 \prod_{i\ne j}
 \frac{\theta(z_i/z_j;q)}{\theta(tz_i/z_j;q)}.
\end{equation}
The power $t^{n(n-1)/2}$ comes from the $n(n-1)/2$ factors
$x_jy_j=t$ in the general identity.

\section{Analytic continuation of the fixed-rank factors}
\label{app:schur-continuation}

This appendix verifies the compatibility used in
Section~\ref{sec:schur-formal-factorization}: the formal product
expressions represent the meromorphic continuations of the fixed-rank
coefficients in \eqref{eq:schur-F-coefficient}, and their radial
expansions are the Taylor expansions of those continued functions.

\subsection{Meromorphic continuation at fixed rank}

Write
$P_1=\sum_{i,j\ge0}c_{ij}A^{-i}B^j$.  In its original convergence
region $|t|>1,|u|<1$, shifting the product index gives
\begin{equation}\label{eq:schur-P-functional}
 P_1(A,B)=(1+A^{-1}+B)P_1(At,Bu).
\end{equation}
Comparing coefficients yields
\begin{equation}\label{eq:schur-P-recurrence}
 (1-t^{-i}u^j)c_{ij}
 =t^{-i}u^{j-1}c_{i,j-1}
  +t^{-(i-1)}u^j c_{i-1,j}.
\end{equation}
An index below zero means that the corresponding coefficient is zero;
the initial coefficient is $c_{00}=1$, and the recurrence is used for
$(i,j)\ne(0,0)$.  It shows directly that every $c_{ij}$ is rational
in $t,u$.

For a fixed $i$, the continued series
$\sum_j c_{ij}B^j$ converges for every finite $B$ when $|u|<1$,
away from its poles in $t$.  To see this, write
$c_{ij}=u^{j(j-1)/2}\widetilde c_{ij}$.  Equation
\eqref{eq:schur-P-recurrence} becomes
\begin{equation}\label{eq:schur-P-growth}
 (1-t^{-i}u^j)\widetilde c_{ij}
 =t^{-i}\widetilde c_{i,j-1}+t^{-(i-1)}u^j \widetilde c_{i-1,j}.
\end{equation}
On a compact parameter set avoiding the poles, its denominators are
bounded away from zero.  For large $j$ they approach $1$ uniformly.
Induction on $i$, followed by induction on $j$, bounds $\widetilde c_{ij}$ by
$C_iK_i^j$ for suitable constants.  The factor
$|u|^{j(j-1)/2}$ then gives convergence for every $B$.
This also gives convergence uniformly on compact parameter sets.
At fixed $i$, the recurrence bounds the pole orders uniformly in $j$:
at a fixed nonzero $t$ and $0<|u|<1$, at most one denominator
$1-t^{-k}u^j$ can vanish for each $1\le k\le i$, and a recurrence
path encounters each such pair at most once.  Thus $[A^{-i}]P_1$ is
entire in $B$ (analytic for every finite $B$) and
meromorphic in nonzero $t$ (analytic except at poles), with possible
pole conditions $t^k=u^j$, $1\le k\le i$, $j\ge0$.
The notation $[A^{-i}]$ means the coefficient of $A^{-i}$.
The statement for $P_2$ follows by interchanging $(t,A^{-1})$ with $(u,B)$.

The purpose of these estimates is to justify extracting Laurent
coefficients in $B$ by contour integration on an annulus after
multiplication by $C_-$. For $C_-$, the required convergence follows
directly from the geometric decay of $u^m$ and $q^m$: for $|u|,|q|<1$
its product converges uniformly on closed annuli in $B\ne0$ and for
bounded $A^{-1}$. Thus each fixed-$A^{-1}$ coefficient of $P_1C_-$ is
holomorphic on such an annulus, away from parameter poles; regularity
at $B=0$ is unnecessary.

It follows that the coefficients of $A^{-a}$ in $P_1C_-$ are ordinary
meromorphic continuations of the first transformed generating
function in \eqref{eq:schur-factorization}.  They agree with that
function when $|t|>1$, $|tu|<1$, and hence continue its fixed-rank
coefficients.  The same argument applies to the coefficients of $B^b$
in $P_2C_+$, beginning in $|u|>1$, $|tu|<1$.
We have continued each fixed-rank coefficient; no assertion that the
whole transformed series converges in the rank variable is needed.

\subsection{Agreement with the radial expansion}

We now check that the formal radial expansion used in
Section~\ref{sec:schur-formal-factorization} agrees with the Taylor
expansion of these continued coefficients.  Fix a particle number $a$
and generic $t_0,u_0$.
For the coefficients $[A^{-i}]P_1$ with $i\le a$, the possible pole
conditions are $t^k=u^j$, with $1\le k\le a$ and $j\ge0$.
After the radial substitution, these poles stay a positive distance
from $\varepsilon=0$. For $j\ne k$ their nonzero moduli are
$(|t_0|^k/|u_0|^j)^{1/(j-k)}$; for fixed $k$ these tend to
$|u_0|^{-1}$ as $j\to\infty$, and the remaining cases form a finite
set. The cases $j=k$ are excluded by genericity. We can therefore
choose a radius $\rho>0$ such that the disk $|\varepsilon|\le\rho$
contains no nonzero pole, and $\rho|t_0|,\rho|u_0|<1$.
Each rational coefficient $c_{ij}$ has a removable singularity at
zero by the nonnegative radial order established in
Section~\ref{sec:schur-formal-factorization}. On the boundary circle
$|\varepsilon|=\rho$, the bound following
\eqref{eq:schur-P-growth} is uniform and is a constant times
$K_i^j(\rho|u_0|)^{j(j-1)/2}$. The maximum-modulus principle, which
bounds an analytic function inside a disk by its boundary values,
extends this bound to the whole disk. Thus
$\sum_j c_{ij}(\varepsilon t_0,\varepsilon u_0)B^j$ converges uniformly
jointly for $|\varepsilon|\le\rho$ and $|B|\le R$, for every finite $R$,
including at $\varepsilon=0$. The tail $C_-$ also converges uniformly
on this disk for bounded $A^{-1}$ and for $B$ in a closed annulus
$0<b_0\le|B|\le b_1<\infty$.
Consequently the radial expansion of each fixed-$A^{-1}$ coefficient of
$P_1C_-$ is its actual Taylor expansion at $\varepsilon=0$.
Interchanging the two species gives the same statement for $P_2C_+$.
The formal product prescription therefore computes the radial Taylor
coefficients of the meromorphically continued fixed-rank functions
specified by \eqref{eq:schur-F-coefficient}.

\phantomsection
\addcontentsline{toc}{section}{References}

\IfFileExists{utphys.bst}{\bibliographystyle{utphys}}{\bibliographystyle{unsrt}} 
\bibliography{ReferenceGGE}


\end{document}